\documentclass[nofootinbib,prd,aps,twocolumn,preprintnumbers,amsmath,amssymb,superscriptaddress]{revtex4}
\usepackage{graphicx} 
\usepackage{amssymb}
\usepackage{amsmath}
\usepackage[colorlinks=true]{hyperref}
\usepackage{ulem}

\begin{document}

\title{Bounce Inflation with Dynamical Dark Energy in Light of DESI DR2}

\author{Xin-zhe Zhang}
\email{zincz@hust.edu.cn}
\affiliation{Department of Astronomy, School of Physics, Huazhong University of Science and Technology,\\Luoyu Road 1037, Wuhan, China}

\author{Hao-Hao Li}
\email{lihaohao@wdu.edu.cn}
\affiliation{Foundational Courses Department, Wuhan Donghu College, \\Wenhua Avenue 301, Wuhan, China}

\author{Taotao Qiu}
\email{qiutt@hust.edu.cn(corresponding author)}
\affiliation{Department of Astronomy, School of Physics, Huazhong University of Science and Technology,\\Luoyu Road 1037, Wuhan, China}

\begin{abstract}
  Recently, the Dark Energy Spectroscopic Instrument Data Release 2 (DESI DR2) suggests that the dark energy in our universe might be evolving, favoring the Chevallier-Polarski-Linder (CPL) parameterization and a lower Hubble constant. In our previous work, it has been reported that cosmological model with the non-singular bounce inflation (BI) scenario and $\Lambda$CDM might alleviate the Hubble tension into 3$\sigma$ confidence. In this paper, we study the cosmological model of BI with a dynamical dark energy. We find that individual consideration of the CPL parameterization and the data \texttt{DESI DR2} tend to larger Hubble constants for both BI and power law (PL) case with cosmic microwave background (CMB) data. Employing BI with combined CPL parameterization and \texttt{DESI DR2}, we obtain the Hubble constant $H_ 0 = 65.2^{ + 1.8}_{ - 2.2} \ \mathrm{km} \cdot \mathrm{s}^{ -1 } \cdot \mathrm{Mpc}^{ -1 }$, which is larger than $H_ 0 = 64.0 \pm 2.1 \ \mathrm{km} \cdot \mathrm{s}^{ -1 } \cdot \mathrm{Mpc}^{ -1 }$ for the PL case. After considering nontrivial weak lensing effect and spatial curvature as well as adding \texttt{Pantheon+}, BI fits 3.1$\sigma$ confidence of $\Lambda$CDM with $w_ 0 = -0.919 \pm 0.038$ and $w_{ \mathrm{a}} = -0.37 \pm 0.12$, and it prefers evolving dark energy than the PL case with $w_ 0 = -0.960 \pm 0.074$ and $w_{ \mathrm{a}} = -0.15^{ +0.28}_{ -0.25}$.
\end{abstract}

\maketitle

\section{Introduction}

Some discrepancies arise between data from early-time and late-time measurements. As a well-known example, the Hubble tension refers to the discrepancy between local measurements based on the cosmic distance ladder and CMB results \cite{DiValentino:2021izs}. Within the framework of the $\Lambda$CDM cosmological model and assuming a power-law form for the primordial power spectrum,
\begin{equation}
    \mathcal{P}_{ \mathcal{R}}( k) = A_{ \mathrm{s}} \left( \frac{ k}{ k_ *} \right)^{ n_{ \mathrm{s}} - 1},
    \label{PL power spectrum}
\end{equation}
the Planck 2018 TT+TE+EE+lowE analysis yields $H_ 0 = 67.27 \pm 0.60 \ \mathrm{km} \cdot \mathrm{s}^{ -1 } \cdot \mathrm{Mpc}^{ -1 }$ \cite{Planck:2018vyg}, the South Pole Telescope (SPT-3G) 2019-2020 data yields $H_ 0 = 66.66 \pm 0.60 \ \mathrm{km} \cdot \mathrm{s}^{ -1 } \cdot \mathrm{Mpc}^{ -1 }$ \cite{SPT-3G:2025bzu}, and the Atacama Cosmology Telescope Data Release 6 (ACT DR6) yields $H_ 0 = 66.11 \pm 0.79 \mathrm{km} \cdot \mathrm{s}^ {-1} \cdot \mathrm{Mpc} ^{-1}$ \cite{ACT:2025fju}. All of these values exhibit a discrepancy with the SH0ES measurements based on the Cepheid-SN Ia of $H_ 0 = 73.04 \pm 1.04 \ \mathrm{km} \cdot \mathrm{s}^{ -1 } \cdot \mathrm{Mpc}^{ -1 }$ \cite{Riess:2021jrx} at more than 5$\sigma$ confidence. The Hubble tension may be indicative of new physics beyond the standard cosmological model, such as modified gravity \cite{Yan:2019gbw, Ren:2022aeo, Boiza:2025xpn}, interacting dark section \cite{Yang:2018euj, Pan:2019gop}, early dark energy \cite{Poulin:2018cxd, Kamionkowski:2022pkx, Chen:2024wqc} or modified primordial epoch \cite{Hazra:2022rdl, Li:2024rgq}. 

A critical aspect in the discussion of Hubble tension is the determination of the observed angle of Baryon Acoustic Oscillation (BAO) sound horizon, in CMB analysis as
\begin{equation}
    \theta_ * \equiv \frac{ r_ *}{ d_{ \mathrm{M}}( z_ *)} = \frac{ \int_{ z_{ *}}^{ +\infty}{ \frac{ c_{ \mathrm{s}}( z)}{ H( z)} dz}}{ \int_{ 0}^{ z_{ *}}{ \frac{ c}{ H( z)} dz}},
    \label{sound horizon of BAO}
\end{equation}
where $\theta_ *$ is the angular scale of the BAO sound horizon at recombination, $r_ *$ is BAO sound horizon at recombination, $d_ M$ is comoving angular diameter distance, $z_ *$ is the redshift at recombination. $c_{ \mathrm{s}}( z) \equiv c / \sqrt{ 3 (1 + \rho_{ \mathrm{b}} / \rho_{ \mathrm{g}})}$ is the sound speed in the photon-baryon fluid with $c$ being the speed of light, $\rho_{ \mathrm{b}}$ as the energy density of baryons and $\rho_{ \mathrm{g}}$ as the energy density of photons. Since $\theta_ *$ is related with the evolution of the universe, the cosmic evolution history is often considered to be modified and must be precisely constrained.

Several additional parameters are also important in fitting for cosmic evolution. One parameter we consider is $S_ 8 \equiv \sigma_ 8 \sqrt{ \Omega_{ \mathrm{m}} / 0.3}$ which represents the amplitude of large-scale structures growth, where $\sigma_ 8$ is the matter fluctuation amplitude parameter on scales of $8 h^{ - 1} \mathrm{Mpc}$, and $\Omega_{ \mathrm{m}}$ is the matter density relative to the critical energy $\rho_{ \mathrm{crit}}$, including the contributions of baryons and cold dark matter. CMB based measurements give $S_ 8 = 0.832 \pm 0.013$ in Planck 2018 \cite{Planck:2018vyg}, $S_ 8 = 0.797 \pm 0.042$ in SPT-3G \cite{SPT-3G:2022hvq}, and $S_ 8 = 0.875 \pm 0.023$ in ACT DR6 \cite{ACT:2025fju}, while local measurements give $S_ 8 = 0.759^{ +0.024}_{ -0.021}$ in Kilo-Degree Survey (KiDS) \cite{KiDS:2020suj}, $S_ 8 = 0.763 \pm 0.009$ in Dark Energy Survey (DES) \cite{DES:2025ucw} and $S_ 8 = 0.769^{ +0.031}_{ -0.034}$ in Hyper Suprime-Cam (HSC) \cite{Li:2023tui}. Recently, with high-redshift calibrations, some re-analyzed results show that, it seems that there is no $S_ 8$ tension between CMB and local measurements, such as $S_ 8 = 0.814^{ + 0.011}_{ - 0.012}$ in KiDS \cite{Stolzner:2025htz, Wright:2025xka}, $S_ 8 = 0.832 ^{ +0.013}_{ -0.017}$ in DES \cite{DarkEnergySurvey:2025bkf} and $S_ 8 = 0.805 \pm 0.018$ in HSC \cite{deJanvry:2025lwc}. However, DESI collaboration still reports small tensions with the results as $S_ 8^{ \text{DESI} \times \text{HSC}}= 0.787 \pm 0.020$, $S_ 8^{ \text{DESI} \times \text{DES}} = 0.791 \pm 0.016$ and $S_ 8^{ \text{DESI} \times \text{KiDS}} = 0.771 \pm 0.017$ \cite{Semenaite:2025ohg}. Although there were also some kind of ``tension'' in $S_8$ measurements, the case is becoming more vague recently.

Another parameter to be considered is weak lensing amplitude $A_{ \mathrm{L}}$ which quantifies the strength of gravitational lensing effects on the CMB. The weak lensing effect smooths the acoustic peaks on CMB angular power spectrum, while the amplitude is normalized to be unity for standard $\Lambda$CDM cosmology \cite{Zaldarriaga:1998ar, Calabrese:2008rt}. In Planck 2018 analysis, $A_{ \mathrm{L}} = 1.180 \pm 0.065$ \cite{Planck:2018vyg} which exceeds unity at approximately 3$\sigma$ significance and is referred as the lensing anomaly. However, recent CMB observations show no significant deviation on $A_{ \mathrm{L}}$, as $A_{ \mathrm{L}} = 1.039 \pm 0.052$ in Planck PR4 \cite{Tristram:2023haj}, $A_{ \mathrm{L}} = 0.972^{ + 0.079}_{ - 0.089}$ in SPT-3G \cite{SPT-3G:2022hvq} and $A_{ \mathrm{L}} = 1.007 \pm 0.057$ in ACT DR6 \cite{ACT:2025fju}.

The spatial curvature also contributes a $\Omega_{ \mathrm{k}}$ term in the expanding rate of the universe and affects the geometric relationship of the measurements of distance. Planck 2018 TT+TE+EE+lowE yields $\Omega_{ \mathrm{k}} = -0.044^{ + 0.033}_{ -0.034}$ tending 1$\sigma$ to a closed universe \cite{Planck:2018vyg}. Meanwhile, $\Omega_{ \mathrm{k}} = -0.012 \pm 0.010$ from Planck DR4 \cite{Tristram:2023haj} and $\Omega_{ \mathrm{k}} = -0.004 \pm 0.010$ from ACT DR6 \cite{ACT:2025fju} both are consistent with a flat universe less than 1$\sigma$ level. SPT-3G yields $\Omega_{ \mathrm{k}} = 0.002^{ +0.015}_{ -0.012}$ is also consistent with a flat universe but with a positive best-fit value. The local observations usually prefer a positive value, such as $\Omega_{ \mathrm{k}} = 0.08^{ +0.16}_{ -0.17}$ from KiDS \cite{Reischke:2025hrt} and $\Omega_{ \mathrm{k}} = 0.055 \pm 0.032$ from DES Y6 combined with DESI DR1 \cite{LozanoTorres:2024xck}. Nevertheless, a flat universe is suggested by the combination of CMB and BAO, like $\Omega_{ \mathrm{k}} = 0.0007 \pm 0.0019$ for Planck 2018 TT+TE+EE+lowE+lensing+BAO \cite{Planck:2018vyg}.

As for the theoretical construction of cosmic evolution history, the Bounce Inflation scenario is proposed to avoid the initial cosmological singularity, in which a contracting stage and a cosmological bounce precede the standard inflationary epoch, see \cite{Qiu:2015nha, Wan:2015hya, Ni:2017jxw, Saha:2024fup, Qiu:2025oop} for concrete models. Observationally, BI could explain the suppression of primordial power spectrum and the hemispherical asymmetry observed in the CMB angular power spectrum at multipoles $\ell < 10$ \cite{Eriksen:2007pc, Hoftuft:2009rq, Planck:2019evm, Sanyal:2024iyv, Piao:2003zm, Cai:2008qb, Liu:2013kea, Xia:2014tda}. Recently we found that a parameterized primordial power spectrum of BI could alleviate the Hubble tension to the 3.2$\sigma$ level, yielding $H_ 0 = 69.38 \pm 0.49 \ \mathrm{km} \cdot \mathrm{s}^{ -1 } \cdot \mathrm{Mpc}^{ -1 }$ in presence of weak lensing effects \cite{Li:2024rgq}. The primordial power spectrum of BI leads to the relative height of acoustic peaks, requiring a revised set of cosmological parameters to achieve a consistent fit. 

On the other hand the observed accelerating expansion of the late universe suggests the existence of dark energy, which exerts negative pressure and is characterized by the equation of state parameter $w < -1 / 3$. The cosmological constant $\Lambda$, as the dark energy in standard $\Lambda$CDM model, is consistent with some observations, such as Planck 2018 \cite{Planck:2018vyg}. However, a recent BAO measurement from DESI DR2, suggests that the dark energy in our universe might be evolving. Phenomenologically, the evolving dark energy described by CPL parameterization is favored by the data \cite{DESI:2025zgx, Chevallier:2000qy, Linder:2002et, Linder:2024rdj}. The CPL parameterization, also known as $w_ 0 w_ a$CDM, provides an effective description of a wide class of physically motivated dark energy models whose equation of state parameter $w(a)$ is given by
\begin{equation}
w( a) = w_ 0 + w_{a}( 1 - a)~.
    \label{CPL parameterization}
\end{equation}
In the limit $w_ 0 = -1$ and $w_{ a} = 0$, the dark energy reduces to the cosmological constant. Since the sound horizon of the BAO $r_ *$ is tightly constrained and observations of the angular scale of the BAO $\theta_ *$ constrain the comoving angular diameter distance $d_{ \mathrm{M}}$. The $w_ 0 w_ a$CDM needs more observations for low redshift objects, like BAO and SN Ia, to constrain the evolution of dark energy. The best-fit parameters of \texttt{DESI DR2} combined with CMB data are $w_{ 0 } = -0.42 \pm 0.21$, $w_{ a } =  -1.75 \pm 0.58$ and the Hubble constant $H_ 0 = 63.6^{ + 1.6 }_{ -2.1 } \ \mathrm{km} \cdot \mathrm{s}^{ -1 } \cdot \mathrm{Mpc}^{ -1 }$\cite{DESI:2025zgx}, which aggravates the Hubble tension. The SN Ia observation \texttt{Pantheon+} does not support dynamical dark energy as $w_ 0 = -0.93 \pm 0.15$ and $w_{ a} = -0.1^{ +0.9}_{ -2.0}$\cite{Brout:2022vxf}. Combining \texttt{DESI DR2} and \texttt{Pantheon+} there are $w_ 0 = -0.888^{ +0.055}_{-0.064}$ and $w_{ a} = -0.17 \pm 0.46$\cite{DESI:2025zgx}.

 In this paper, we try to invesigate that when combined with the dark energy described by CPL parameterization, whether BI can remain consistent with both early-time and late-time observations. For the numerically analysis of cosmological models, we employ the Markov Chain Monte Carlo sampler \texttt{MontePython} \cite{Audren:2012wb, Brinckmann:2018cvx}, interfaced with Einstein-Boltzmann equation solver \texttt{Class} \cite{Blas:2011rf}. CMB data used in this analysis are \texttt{Planck 2018}: \texttt{Planck\_highl\_TTTEEE} + \texttt{Planck\_lowl\_EE} + \texttt{Planck\_lowl\_TT}\footnote{\url{https://pla.esac.esa.int/pla/\#cosmology}}, \texttt{SPT3G-Y1} \footnote{\url{https://github.com/SouthPoleTelescope/spt3g_y1_dist}} and \texttt{ACT DR4} \footnote{\url{https://github.com/ACTCollaboration/pyactlike}}.
These three data are combined and referred to as \texttt{PSA} in the subsequent analysis.
There are also \texttt{DESI DR2} for the BAO observation and \texttt{Pantheon+} for the SN Ia observation. We use \texttt{GetDist} \cite{Lewis:2019xzd} to post-process the MCMC chains. Convergence of these MCMC chains is assessed using the Gelman-Rubin criterion with $R - 1 < 0.001$.

The organization of this paper is as follows: in SEC. \ref{Section: Bounce Inflation and Cosmological tensions}, we review the introduction of BI for alleviating cosmological tensions; in SEC. \ref{Section: The role of DESI DR2 and CPL}, we investigate the role of \texttt{DESI DR2} in $\Lambda$CDM and CPL parameterization without large-scale structure constraints; in SEC. \ref{Section: Bounce inflation and CPL}, we present the numerical analysis of BI and PL using \texttt{PSA}, \texttt{DESI DR2} and \texttt{Pantheon+}, within the CPL parameterization framework; finally, we present our conclusions and discussion in SEC.\ref{Section: Conclusion}.

\section{Bounce Inflation and Cosmological Tensions}
\label{Section: Bounce Inflation and Cosmological tensions}

\begin{figure}
    \centering
    \includegraphics[width=0.9\linewidth]{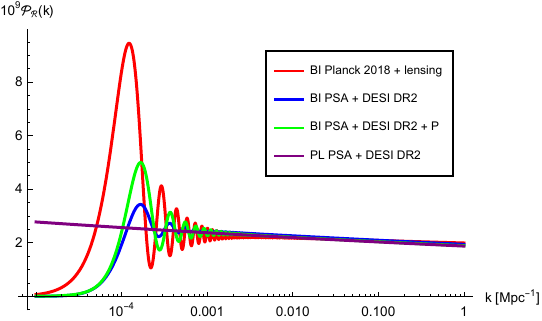}
    \caption{The best-fit results of the BI and PL power spectrum using \texttt{Planck 2018} + \texttt{Planck\_lensing}, \texttt{PSA} + \texttt{DESI DR2} and \texttt{PSA} + \texttt{DESI DR2} + \texttt{P} (\texttt{P} stands for \texttt{Pantheon+}). The oscillating behavior of BI power spectrum is suppressed when fitting to \texttt{PSA} + \texttt{DESI DR2} data and has a lower index $n_{ \mathrm{s}} \approx 0.97$ than the BI scenario using \texttt{Planck 2018} + \texttt{Planck\_lensing}, similar to the spectral index in power law case with \texttt{PSA} + \texttt{DESI DR2}.}
    \label{fig:BI power spectrum}
\end{figure}

Firstly, we review the parameterization of the BI scenario and its associated numerical analysis, see \cite{Li:2024rgq} for more details. The BI scenario has a contraction phase and a cosmological bounce before the inflationary phase. In this parameterization of BI, the slow-roll parameter $\epsilon \equiv - \dot{ H} / H^ 2$ at contraction and expansion are assumed as constants. It allows an analytic formulation of the scale factor as \cite{Ni:2017jxw}
\begin{equation}
    a( \eta) = \begin{cases}
        a_{ \mathrm{c}}( \tilde{ \eta}_{ \mathrm{B}_ -} - \eta)^{ \frac{ 1}{ \epsilon_{ \mathrm{c}} - 1}} &\eta < \eta_{ \mathrm{B}_ -}\\
        a_{ \mathrm{B}} \left[ 1 + \frac{ \alpha}{ 2}( \eta - \eta_{ \mathrm{B}})^ 2 \right] &\eta_{ \mathrm{B}_ -} \leq \eta \leq \eta_{ \mathrm{B}_ +}\\
        a_{ \mathrm{e}}( \tilde{ \eta}_{ \mathrm{B}_+} - \eta)^{ \frac{ 1}{ \epsilon_{ \mathrm{e}} - 1}} &\eta > \eta_{ \mathrm{B}_+}
    \end{cases},
    \label{parameterization of BI}
\end{equation}
where $a(\eta)$ is the scale factor as a function of conformal time $\eta$, $a_{ \mathrm{c}}$, $a_{ \mathrm{B}}$ and $a_{\mathrm{e}}$ are the scale factor at the bounce beginning $\eta_{ \mathrm{B}_ -}$, the bounce point $\eta_{ \mathrm{B}}$ and the bounce ending $\eta_{ \mathrm{B}_ +}$ correspondingly. $\tilde{ \eta}_{ \mathrm{B}_ -} \equiv \eta_{ \mathrm{B}_ -} - [( \epsilon_{ \mathrm{c}} - 1) \mathcal{H}_{ \mathrm{c}}]^{ -1}$, where $\mathcal{H}_{ \mathrm{c}}$ as the co-moving Hubble parameter at $\eta_{ \mathrm{B}_ -}$, $\tilde{ \eta}_{ \mathrm{B}_ +} \equiv \eta_{ \mathrm{B}_ +} - [( \epsilon_{ \mathrm{e}} - 1) \mathcal{H}_{ \mathrm{e}}]^{ -1}$ with $\mathcal{H}_{ \mathrm{e}}$ as the co-moving Hubble parameter at $\eta_{ \mathrm{B}_ +}$. $\epsilon_{ \mathrm{c}}$ and $\epsilon_{ \mathrm{e}}$ are slow-roll parameters of contraction and expansion. Scalar perturbations in BI are governed by the Mukhanov-Sasaki equation:
\begin{equation}
    u_{ k}^{ \prime \prime} + \left( k^{ 2} c_{ \mathrm{s}}^ 2 - \frac{ z^{ \prime \prime}}{ z} \right) u_{ k}= 0,
    \label{M-S equation}
\end{equation}
where $u_{ k}$ is the Fourier mode of Mukhanov-Sasaki variable $u \equiv z \zeta$ with $\zeta$ as co-moving curvature perturbation, $k$ is wavenumber, $c_{\mathrm{s}}$ is the sound speed and $z \equiv a \sqrt{ Q} / c_{ \mathrm{s}}$ with $Q$ derived from second-order perturbed action \cite{Qiu:2015nha}. The primordial scalar power spectrum is obtained by solving Eq. \eqref{M-S equation} using the scale factor given in \eqref{parameterization of BI} in each stage with matching conditions at $\eta_{ \mathrm{B}_ -}$ and $\eta_{ \mathrm{B}_ +}$, and the result is \cite{Li:2024rgq}
\begin{equation}
\begin{split}
    P_{ \mathcal{R}}( k) &= \frac{ H_{ \mathrm{e}}^{ 2}}{ 8 \pi^{ 2} M_{ \mathrm{P}}^{ 2} \epsilon_{ \mathrm{e}}} \left( \frac{ k}{ k_{ *}} \right)^{ 3 - 2 \nu_{ \mathrm{e}}} \left| C_{ 1} - C_{ 2} \right|^{ 2}\\
    &= A_{ \mathrm{s}} \left( \frac{ k}{ k_ *} \right)^{ n_{ \mathrm{s}} -1} \left| C_{ 1} - C_{ 2} \right|^{ 2},
\end{split}
\label{power spectrum of BI}
\end{equation}
where $M_{ \mathrm{P}}$ is the reduced Planck mass, $\nu_{ \mathrm{e}} \equiv (\epsilon_{ \mathrm{e}} - 3) / [ 2( \epsilon_{ \mathrm{e}} - 1)]$. The coefficients $C_{ 1}$ and $C_{ 2}$ with $c_{ \mathrm{s}} = 1$ are
\begin{widetext}
\begin{equation}
	\begin{split}
		C_ 1 &= - i e^{ i \nu_{ \mathrm{c}1} \pi} \sqrt{ \frac{ \tilde{ \mathcal{ H}}_{ \mathrm{e}}}{ \tilde{ \mathcal{ H}}_{ \mathrm{c}}}} \pi^{ 3 / 2} \sin{(l \Delta \eta_{ \mathrm{B}})}\bigg\lbrace k \bigg[ H^{ (1)} _{ \nu_{ \mathrm{c}} + 1} \bigg( \frac{ k}{ \tilde{ \mathcal{ H}}_{ \mathrm{c}}} \bigg) - H^{ (1)} _{ \nu_{ \mathrm{c}} - 1} \bigg( \frac{ k}{ \tilde{ \mathcal{ H}}_{ \mathrm{c}}} \bigg) \bigg]\bigg[ k \bigg( H^{ (2)}_{ \nu_{ \mathrm{e}} + 1} \bigg( \frac{ k}{ \tilde{ \mathcal{ H}}_{ \mathrm{e}}} \bigg)\\
        &- H^{ (2)}_{ \nu_{ \mathrm{e}} - 1} \bigg( \frac{ k}{ \tilde{ \mathcal{ H}}_{ \mathrm{e}}} \bigg) \bigg) - \bigg( \tilde{ \mathcal{ H}}_{ \mathrm{e}} + 2 l \cot{(l \Delta \eta_{ \mathrm{B}})} \bigg) H^{ (2)}_{ \nu_{ \mathrm{e}}} \bigg( \frac{ k}{ \tilde{ \mathcal{ H}}_{ \mathrm{e}}} \bigg) \bigg] + H^{ (1)}_{ \nu_{ \mathrm{c}}} \bigg( \frac{ k}{ \tilde{ \mathcal{ H}}_{ \mathrm{c}}} \bigg) \bigg[ \tilde{ \mathcal{ H}}_{ \mathrm{e}} \tilde{ \mathcal{ H}}_{ \mathrm{c}} + 4 l^ 2  \\
		&+ 2 \big( \tilde{ \mathcal{ H}}_{ \mathrm{c}} - \tilde{ \mathcal{ H}}_{ \mathrm{e}} \big) l \cot{(l \Delta \eta_{ \mathrm{B}})} \bigg] H^{ (2)}_{ \nu_{ \mathrm{e}}} \bigg( \frac{ k}{ \tilde{ \mathcal{ H}}_{ \mathrm{e}}} \bigg) + k \bigg( 2 l \cot{(l \Delta \eta_{ \mathrm{B}})} - \tilde{ \mathcal{ H}}_{ \mathrm{c}} \bigg) \bigg( H^{ (2)}_{ \nu_{ \mathrm{e}} + 1} \bigg( \frac{ k}{ \tilde{ \mathcal{ H}}_{ \mathrm{e}}} \bigg) \\
		&- H^{ (2)}_{ \nu_{ \mathrm{e}} - 1} \bigg( \frac{ k}{ \tilde{ \mathcal{ H}}_{ \mathrm{e}}} \bigg) \bigg) \bigg\rbrace \bigg/ \bigg[ 16 l \tilde{ \mathcal{ H}}_{ \mathrm{e}} + 8 l k \pi \bigg( J_{ \nu_{ \mathrm{e}}} \bigg( \frac{ k}{ \tilde{ \mathcal{ H}}_{ \mathrm{e}}} \bigg) Y_{ \nu_{ \mathrm{e}} - 1} \bigg( \frac{ k}{ \tilde{ \mathcal{ H}}_{ \mathrm{e}}} \bigg) - J_{ \nu_{ \mathrm{e}} - 1} \bigg( \frac{ k}{ \tilde{ \mathcal{ H}}_{ \mathrm{e}}} \bigg) Y_{ \nu_{ \mathrm{e}}} \bigg( \frac{ k}{ \tilde{ \mathcal{ H}}_{ \mathrm{e}}} \bigg) \bigg) \bigg]
	\end{split},
\end{equation}
\end{widetext}
and
\begin{widetext}
\begin{equation}
	\begin{split}
		C_ 2 &= i e^{ i \nu_{ c1} \pi} \sqrt{ \frac{ \tilde{ \mathcal{ H}}_e}{ \tilde{ \mathcal{ H}}_{ \mathrm{c}}}} \pi^{ 3 / 2} \sin{(l \Delta \eta_{ \mathrm{B}})} \bigg\lbrace k \bigg[ H^{ (1)} _{ \nu_{ \mathrm{c}} + 1} \bigg( \frac{ k}{ \tilde{ \mathcal{ H}}_{ \mathrm{c}}} \bigg) - H^{ (1)} _{ \nu_{ \mathrm{c}} - 1} \bigg( \frac{ k}{ \tilde{ \mathcal{ H}}_{ \mathrm{c}}} \bigg) \bigg] \bigg[ k \bigg( H^{ (1)}_{ \nu_{ \mathrm{e}} + 1} \bigg( \frac{ k}{ \tilde{ \mathcal{ H}}_{ \mathrm{e}}} \bigg) \\
        &- H^{ (1)}_{ \nu_{ \mathrm{e}} - 1} \bigg( \frac{ k}{ \tilde{ \mathcal{ H}}_{ \mathrm{e}}} \bigg) \bigg) - \bigg( \tilde{ \mathcal{ H}}_{ \mathrm{e}} + 2 l \cot{(l \Delta \eta_{ \mathrm{B}})} \bigg) H^{ (1)}_{ \nu_{ \mathrm{e}}} \bigg( \frac{ k}{ \tilde{ \mathcal{ H}}_{ \mathrm{e}}} \bigg) \bigg] + H^{ (1)}_{ \nu_{ \mathrm{c}}} \bigg( \frac{ k}{ \tilde{ \mathcal{ H}}_{ \mathrm{c}}} \bigg) \bigg[ \tilde{ \mathcal{ H}}_{ \mathrm{e}} \tilde{ \mathcal{ H}}_{ \mathrm{c}} + 4 l^ 2 \\
		&+ 2 \big( \tilde{ \mathcal{ H}}_{ \mathrm{c}} - \tilde{ \mathcal{ H}}_{ \mathrm{e}} \big) l \cot{(l \Delta \eta_{ \mathrm{B}})} \bigg] H^{ (1)}_{ \nu_{ \mathrm{e}}} \bigg( \frac{ k}{ \tilde{ \mathcal{ H}}_{ \mathrm{e}}} \bigg) + k \bigg( 2 l \cot{(l \Delta \eta_{ \mathrm{B}})} - \tilde{ \mathcal{ H}}_{ \mathrm{c}} \bigg) \bigg( H^{ (1)}_{ \nu_{ \mathrm{e}} + 1} \bigg( \frac{ k}{ \tilde{ \mathcal{ H}}_{ \mathrm{e}}} \bigg)\\
		& - H^{ (1)}_{ \nu_{ \mathrm{e}} - 1} \bigg( \frac{ k}{ \tilde{ \mathcal{ H}}_{ \mathrm{e}}} \bigg) \bigg) \bigg\rbrace \bigg/ \bigg[ 16 l \tilde{ \mathcal{ H}}_{ \mathrm{e}} + 8 l k \pi \bigg( J_{ \nu_{ \mathrm{e}}} \bigg( \frac{ k}{ \tilde{ \mathcal{ H}}_{ \mathrm{e}}} \bigg) Y_{ \nu_{ \mathrm{e}} - 1} \bigg( \frac{ k}{ \tilde{ \mathcal{ H}}_{ \mathrm{e}}} \bigg) - J_{ \nu_{ \mathrm{e}} - 1} \bigg( \frac{ k}{ \tilde{ \mathcal{ H}}_{ \mathrm{e}}} \bigg) Y_{ \nu_{ \mathrm{e}}} \bigg( \frac{ k}{ \tilde{ \mathcal{ H}}_{ \mathrm{e}}} \bigg) \bigg) \bigg]  
	\end{split},
\end{equation}
\end{widetext}
where $\tilde{H}_{ \mathrm{c}} \equiv( 1 - \epsilon_{ \mathrm{c}}) H_{ \mathrm{c}}$ with $H_{ \mathrm{c}}$ as Hubble parameter at $\eta_{ \mathrm{B}_ -}$, $\tilde{H}_{ \mathrm{e}} \equiv( 1 - \epsilon_{ \mathrm{e}}) H_{ \mathrm{e}}$ with $H_{ \mathrm{e}}$ as Hubble parameter at $\eta_{ \mathrm{B}_ +}$, $l \equiv \sqrt{ c_{ \mathrm{s}}^{ 2} k^{ 2} - ( \alpha - \chi) a_{ \mathrm{B}}^{ 2}}$, $\Delta \eta_{ \mathrm{B}} \equiv \eta_{ \mathrm{B}_ +} - \eta_{ \mathrm{B}_ -}$ and $\nu_{ \mathrm{c}} \equiv ( \epsilon_{ \mathrm{c}} - 3) / [ 2 ( \epsilon_{ \mathrm{c}} - 1)]$. $H_ a ^{ (1)}$ and $H_ a ^{ (2)}$ are the first and second Hankel functions of order $a$, $J_{ a}$ and $Y_ a$ are the first and second Bessel function of order $a$, respectively \cite{Li:2024rgq}. To avoid contraction anisotropy, $\epsilon_{ \mathrm{c}}$ should be no less than 3. The power spectrum is shown in FIG. \ref{fig:BI power spectrum} as samples.

Due to the discrepancy between BI and PL power spectrum, cosmological parameters have different best-fits for BI compared to the numerical analysis result of Planck 2018. With the data \texttt{Planck 2018} + \texttt{Planck\_lensing}, the best-fit value of the Hubble constant in BI is $H_ 0 = 69.38 \pm 0.49 \ \mathrm{km} \cdot \mathrm{s}^{ -1 } \cdot \mathrm{Mpc}^{ -1 }$ when allowing for $A_{ \mathrm{L}} = 1.128 \pm 0.038$ as previously noted, and $H_ 0 = 68.60^{ + 0.40}_{ - 0.45} \ \mathrm{km} \cdot \mathrm{s}^{ -1 } \cdot \mathrm{Mpc}^{ -1 }$ when $A_{ \mathrm{L}} = 1$ is fixed. What's more, a larger optical depth of reionization $\tau_{ \mathrm{reio}} = 0.0616 \pm 0.0072$ and a larger spectral index $n_{ \mathrm{s}} = 0.9802 \pm 0.0043$ of BI with $A_{ \mathrm{L}} = 1$ are obtained, comparing to the result of Planck 2018, which are $\tau_{ \mathrm{reio}} = 0.0544 \pm 0.0073$ and $n_{ \mathrm{s}} =  0.9649 \pm 0.0042$. The suppression and oscillation at small $k$ from BI power spectrum, as shown in FIG. \ref{fig:BI power spectrum}, affect the angular power spectrum of CMB, and are compensated by adjusting other cosmological parameters, especially a larger $\tau_{ \mathrm{reio}}$ and a larger $n_{ \mathrm{s}}$. Thus the degeneracy between cosmological parameters, as discussed in \cite{Li:2025nnk}, will contribute to a slightly larger $H_ 0$, and alleviate Hubble tension to 3.2$\sigma$ level.

\section{Separating \texttt{DESI DR2} and CPL parameterization}
\label{Section: The role of DESI DR2 and CPL}

The recent data of BAO observation from DESI supports an evolving dark energy. The CPL parameterization \eqref{CPL parameterization} provides a phenomenological description of an evolving dark energy model. In CPL parametrization, since
\begin{equation}
    \frac{ \rho_{ \mathrm{DE}}( a)}{ \rho_{ \mathrm{DE},0}} = a^{ - 3( 1 + w_ 0 + w_ {a})} e^{ - 3 w_{a}( 1 - a)}~,
    \label{state of equation of motion}
\end{equation}
a larger $w_ 0 $ or a lower energy density of dark energy $\rho_{ \mathrm{DE}}$ at low redshift about $z < 0.5$ results in a lower $H_ 0$, so there should be a larger $\rho_{ \mathrm{DE}}$ at higher redshift about $z > 0.5$ with a less $w_ a$ to obtain the same value of $d_{ \mathrm{M}}$. Conversely, a less $w_ 0$ or a larger $\rho_{ \mathrm{DE}}$ constitutes a set with a larger $w_ a$ and a larger $H_ 0$. The BAO observations at different redshift is needed to constrain the evolution of dark energy at late universe.

\begin{figure*}
    \centering
    \includegraphics[width=0.9\linewidth]{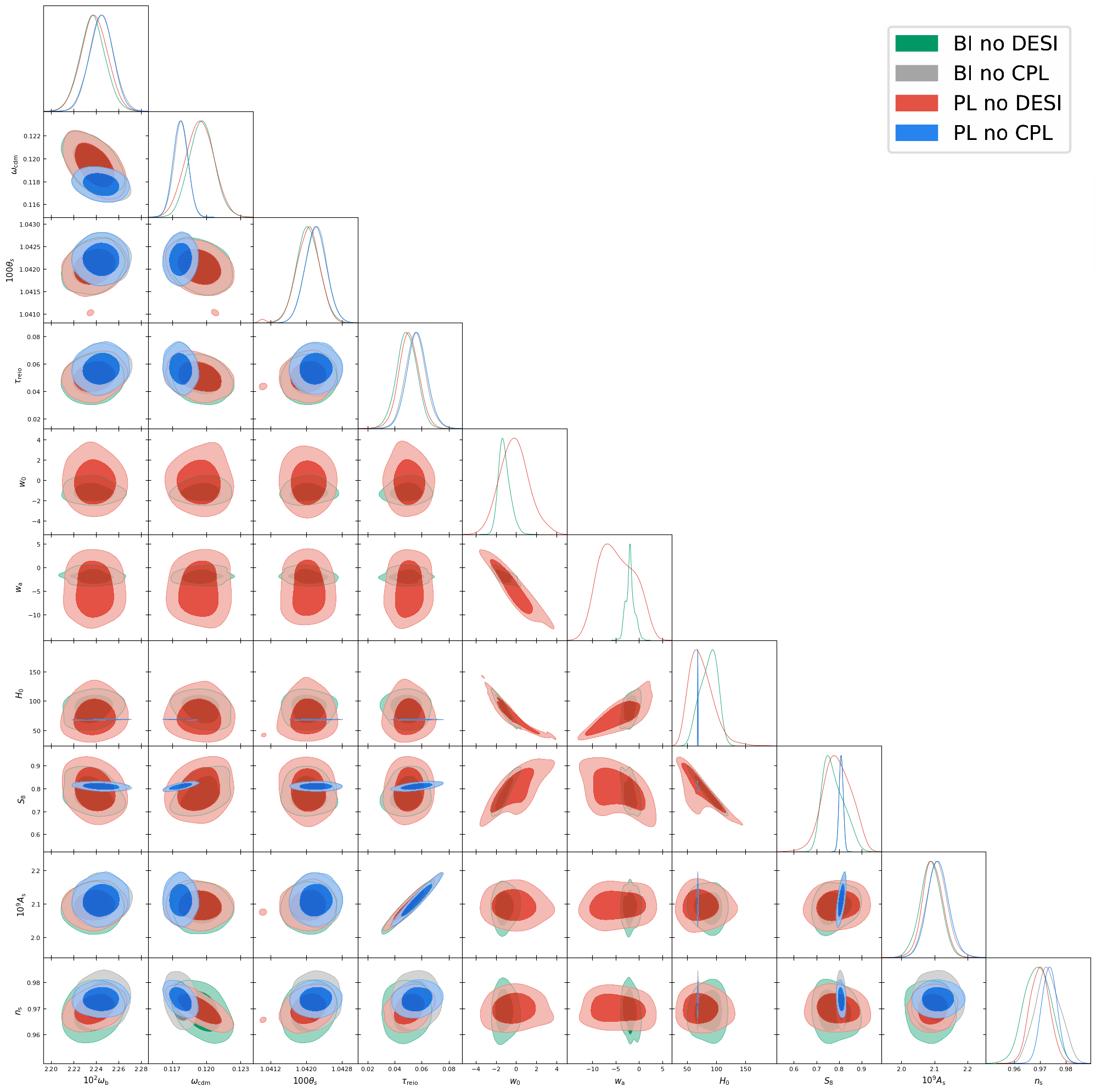}
    \caption{Correlations of cosmological parameters of BI and PL with \texttt{PSA} + \texttt{DESI DR2} based on $\Lambda$CDM or \texttt{PSA} based on CPL parameterization. The \texttt{DESI DR2} favors a larger $r_ *$ and thus a larger $H_ 0$ than CMB results. On the other hand, because \texttt{PSA} could not constrain the CPL parameters, cases of \texttt{PSA} based on CPL parameterization have larger $H_ 0$ and less $S_ 8$.}
    \label{fig:DESI and CPL cosmo}
\end{figure*}

\begin{table*}[!t]
\centering
\begin{tabular} { | l | c c | c c | }
\hline

Data & \multicolumn{2}{|c|}{PSA} & \multicolumn{2}{|c|}{PSA + DESI}\\

\hline

Model &  BI + CPL &  PL + CPL &  BI + $\Lambda$CDM &  PL + $\Lambda$CDM\\
 


\hline
{\boldmath$10^{2}\omega{}_{\mathrm{b} }$} & $2.237\pm 0.011            $ & $2.239\pm 0.012            $ & $2.245\pm 0.011            $ & $2.244\pm 0.010            $\\

{\boldmath$\omega{}_{\mathrm{cdm} }$} & $0.1196\pm 0.0011          $ & $0.1194\pm 0.0012          $ & $0.11778\pm 0.00062        $ & $0.11774\pm 0.00064        $\\

{\boldmath$100\theta{}_{ \mathrm{s} }$} & $1.04204\pm 0.00026        $ & $1.04203\pm 0.00028        $ & $1.04222\pm 0.00024        $ & $1.04221\pm 0.00024        $\\

{\boldmath$\tau{}_{\mathrm{reio} }$} & $0.0490\pm 0.0076          $ & $0.0505\pm 0.0076          $ & $0.0556\pm 0.0076          $ & $0.0566\pm 0.0077          $\\

{\boldmath$w_ 0      $} & $-1.17^{+0.46}_{-0.67}     $ & $-0.2\pm 1.4               $ & $-1$ & $-1$ \\

{\boldmath$w_{ \mathrm{a} }      $} & $-1.91^{+0.68}_{-1.2}      $ & $-4.7^{+3.5}_{-4.4}        $ & 0 & 0 \\

\hline

{\boldmath$H_0            $} & $87^{+20}_{-10}            $ & $75^{+20}_{-20}            $ & $68.37\pm 0.28             $ & $68.37\pm 0.27             $\\

{\boldmath$S_{8 }         $} & $0.776^{+0.036}_{-0.058}   $ & $0.793\pm 0.062            $ & $0.8093\pm 0.0091          $ & $0.8098\pm 0.0091          $\\

{\boldmath$10^{9}A_{\mathrm{s} }$} & $2.089\pm 0.032            $ & $2.094\pm 0.031            $ & $2.108\pm 0.032            $ & $2.111\pm 0.033            $\\

{\boldmath$n_{\mathrm{s} }  $} & $0.9690\pm 0.0051          $ & $0.9699\pm 0.0038          $ & $0.9730\pm 0.0048          $ & $0.9736\pm 0.0030          $\\
\hline

Label & BI no DESI & PL no DESI & BI no CPL & PL no CPL\\

\hline
\end{tabular}
\caption{Best-fits and 1$\sigma$ uncertainties of BI and PL without \texttt{DESI} or without CPL parameterization. These cases are labeled as ``BI no DESI'', ``PL no DESI'', ``BI no CPL'' and ``PL no CPL'' in order as FIG. \ref{fig:DESI and CPL cosmo}. $\theta_{ \mathrm{s}}$ is slightly different from $\theta_ *$. In \texttt{CLASS} the redshift to calculate $\theta_{ \mathrm{s}}$ is chosen at the decoupling time given by maximum of visibility function $\beta( z) \equiv - e^{ - \tau( z)} d\tau / dz$, which is very close to $z_ *$.}
\label{tab:DESI and CPL cosmo}
\end{table*}

In order to get an insight into the effects from CPL parametrization and the \texttt{DESI DR2} data, we make constraints on the CPL parameterization model using \texttt{PSA} alone and on the $\Lambda$CDM using \texttt{PSA} + \texttt{DESI DR2} individually, as shown in FIG. \ref{fig:DESI and CPL cosmo} and TAB. \ref{tab:DESI and CPL cosmo}. For the constraint on the CPL parameterization using \texttt{PSA} alone, showing in TAB. \ref{tab:DESI and CPL cosmo} as ``BI no DESI'' with the power spectrum of BI and `` PL no DESI'' with the PL, we get $\omega_{ \mathrm{b}} \equiv \Omega_{ \mathrm{b}} h^ 2$ and $\omega_{ \mathrm{cdm}} \equiv \Omega_{ \mathrm{cdm}} h^ 2$ which have no deviation from the results of Planck 2018, where $h \equiv H_ 0 / ( 100 \ \mathrm{km} \cdot \mathrm{s}^{ -1 } \cdot \mathrm{Mpc}^{ -1 })$ is the reduced Hubble constant. However, since $d_M$ is primarily determined by $H(z)$ at low redshift, $H_ 0$ is much larger with huge errors as $H_ 0 = 87^{ + 20}_{ - 10} \ \mathrm{km} \cdot \mathrm{s}^{ -1 } \cdot \mathrm{Mpc}^{ -1 }$ for ``BI no DESI'' and $H_ 0 = 75^{ + 20}_{ - 20} \ \mathrm{km} \cdot \mathrm{s}^{ -1 } \cdot \mathrm{Mpc}^{ -1 }$ for ``PL no DESI''.  ``BI no DESI'' has a less $w_ 0$ and a larger $H_ 0$ than ``PL no DESI''. This indicates that \texttt{PSA} only provides weak constraints on CPL parameterization, allowing a broad range of $H_ 0$. Moreover, $S_ 8 = 0.776^{ + 0.036}_{ - 0.058}$ in ``BI no DESI'' is less than $S_ 8 = 0.793 \pm 0.062$ in ``PL no DESI''. Nevertheless, there is a clear tendency toward a larger $H_ 0$ and a less $S_ 8$. These results are consistent with Planck 2018 \cite{Planck:2018vyg}.

For the $\Lambda$CDM model constrained with \texttt{PSA} + \texttt{DESI DR2}, labeled  as ``BI no CPL'' with BI scenario and ``PL no CPL'' with PL in TAB. \ref{tab:DESI and CPL cosmo}, cosmological parameters are in well agreement with each other within 1$\sigma$ confidence. Compared to Planck 2018 results, $10^ 2 \omega_{ \mathrm{b}} \approx 2.244$ is slightly larger, $\omega_{ \mathrm{cdm}} \approx 0.1177$ is less, $\tau_{ \mathrm{reio}} \approx 0.56$ is larger and $n_{ \mathrm{s}} \approx 0.973$ is slightly larger in both cases. Both cases yield a larger value of $100 \theta_{ \mathrm{s}} \approx 1.0422$ than ``no DESI'' cases. $\omega_{ \mathrm{m}} = \omega_{ \mathrm{b}} + \omega_{ \mathrm{cdm}}$ is reduced comparing to Planck 2018, leading to a slightly larger BAO sound horizon as $r_ * \approx 145.04\ \mathrm{Mpc}$ than $r_ * = 144.43 \pm 0.26\ \mathrm{Mpc}$ in Planck 2018 \cite{Planck:2018vyg}. At the same time, numerical analyzes of both cases agree with a larger $\rho_{ \mathrm{DE},0}$, thus a larger $H_ 0 (\approx 68.37)$ and a less $S_ 8 (\approx 0.81)$ than Planck 2018. These results imply that \texttt{DESI DR2} prefers a larger $\theta_ *$ than \texttt{PSA}, which is consistent with the results of DESI Collaborations \cite{DESI:2025zgx}.

\section{Bounce inflation with the CPL dark energy}
\label{Section: Bounce inflation and CPL}

\begin{table*}[!t]
    \centering
    \begin{tabular} { | l | c c c c c | }
\hline

 Parameter &  BI &  BI + $A_{ \mathrm{L}}$ &  BI + $\Omega_{ \mathrm{k}}$ &  BI + $A_{ \mathrm{L}}\&\Omega_{ \mathrm{k}}$   &   BI + $A_{\mathrm{L}}\&\Omega_{\mathrm{k}}$ (with \texttt{Pantheon+})   \\
\hline
{\boldmath$10^{2}\omega{}_{\mathrm{b} }$} & $2.236\pm 0.011            $ & $2.237\pm 0.012            $ & $2.233\pm 0.012            $ & $2.238\pm 0.012            $ & $2.235\pm 0.012            $ \\

{\boldmath$\omega{}_{\mathrm{cdm} }$} & $0.11964\pm 0.00097        $ & $0.1195\pm 0.0011          $ & $0.1202\pm 0.0012          $ & $0.1193\pm 0.0012          $ & $0.12029\pm 0.00099        $ \\

{\boldmath$10^ 2\theta{}_{ \mathrm{s} }$} & $1.04208\pm 0.00024        $ & $1.04204\pm 0.00024        $ & $1.04199\pm 0.00025        $ & $1.04204\pm 0.00026        $ & $1.04196\pm 0.00024        $ \\

{\boldmath$\tau{}_{\mathrm{reio} }$} & $0.0511\pm 0.0078          $ & $0.0479\pm 0.0083          $ & $0.0504\pm 0.0076          $ & $0.0484\pm 0.0085          $ & $0.0476\pm 0.0083          $ \\

{\boldmath$w_{0}      $} & $-0.44\pm 0.14             $ & $-0.45\pm 0.24             $ & $-0.46^{+0.23}_{-0.28}     $ & $-0.61\pm 0.21             $ & $-0.919\pm 0.038           $ \\

{\boldmath$w_{ \mathrm{a}}      $} & $-1.63\pm 0.41             $ & $-1.60^{+0.71}_{-0.59}     $ & $-1.62^{+0.84}_{-0.63}     $ & $-1.11\pm 0.60             $ & $-0.37\pm 0.12             $ \\

{\boldmath$\epsilon_{ \mathrm{c}}        $} & $3.61^{+0.31}_{-0.50}      $ & $4.7^{+1.6}_{-1.8}         $ & $9.6^{+2.8}_{-6.4}         $ & $4.8^{+1.1}_{-1.7}         $ & $3.29\pm 0.18              $ \\

{\boldmath$10^{2}\epsilon_{ \mathrm{e}} $} & $1.48^{+0.25}_{-0.28}      $ & $1.51\pm 0.16              $ & $1.59^{+0.16}_{-0.18}      $ & $1.46\pm 0.16              $ & $1.656^{+0.082}_{-0.059}   $ \\

{\boldmath$10^{4}\mathrm{H}_{ \mathrm{c}} $} & $-2.3\pm 1.2               $ & $-2.8\pm 1.4               $ & $-2.20^{+2.1}_{-0.47}      $ & $-1.7^{+1.6}_{-2.3}        $ & $-3.50^{+0.25}_{-0.34}     $ \\

{\boldmath$10^{5}\mathrm{H}_{ \mathrm{e}} $} & $5.57\pm 0.39              $ & $5.62\pm 0.31              $ & $5.78\pm 0.31              $ & $5.53^{+0.34}_{-0.30}      $ & $5.89^{+0.15}_{-0.13}      $ \\

{\boldmath$\Delta \eta_{ \mathrm{B}}        $} & $1.53^{+0.65}_{-0.74}      $ & $18.1^{+4.4}_{-2.6}        $ & $19^{+14}_{-19}            $ & $1.9^{+1.2}_{-1.7}         $ & $1.30^{+0.35}_{-0.26}      $ \\

{\boldmath$A_{\mathrm{L} }         $} &   $1$                            & $1.040\pm 0.047            $ &              $1$                &$1.047\pm 0.048            $ & $1.050^{+0.035}_{-0.039}   $ \\

{\boldmath$\Omega{}_{\mathrm{k} }  $} &      $0$                         &               $0$               & $0.0006^{+0.0016}_{-0.0014}$ &$0.0006\pm 0.0015          $ & $0.00172^{+0.00085}_{-0.0011}$ \\

\hline

{\boldmath$\Omega_{ \mathrm{DE},0}  $} & $0.649\pm 0.016            $ & $0.650^{+0.024}_{-0.022}   $ & $0.649^{+0.027}_{-0.021}   $ & $0.664\pm 0.020            $ & $0.6942\pm 0.0057          $ \\

{\boldmath$H_0            $} & $63.8^{+1.3}_{-1.6}        $ & $63.9\pm 2.1               $ & $64.1\pm 2.2               $ & $65.2^{+1.8}_{-2.2}        $ & $68.66^{+0.63}_{-0.73}     $ \\

{\boldmath$S_{8 }         $} & $0.846\pm 0.013            $ & $0.841\pm 0.018            $ & $0.849\pm 0.016            $ & $0.832\pm 0.017            $ & $0.8234\pm 0.0098          $ \\

{\boldmath$10^{9}A_{\mathrm{s} }$} & $2.098\pm 0.033            $ & $2.083\pm 0.036            $ & $2.098\pm 0.033            $ & $2.085\pm 0.036            $ & $2.088\pm 0.034            $ \\

{\boldmath$n_{\mathrm{s} }  $} & $0.9700^{+0.0053}_{-0.0069}$ & $0.9693\pm 0.0045          $ & $0.9677\pm 0.0047          $ & $0.9701\pm 0.0051          $ & $0.9663^{+0.0032}_{-0.0052}$ \\
\hline

$\chi^ 2_{ \mathrm{min}}$ & 2490.96 & 2489.11 & 2487.75 & 2490.32 & 3006.71 \\

\hline

\end{tabular}

\begin{tabular} { | l | c c c c c | }
\hline

 Parameter &  PL &  PL + $A_{ \mathrm{L}}$ &  PL + $\Omega_{ \mathrm{k}}$ &  PL + $A_{ \mathrm{L}}\&\Omega_{ \mathrm{k}}$   &   PL + $A_{\mathrm{L}}\&\Omega_{\mathrm{k}}$ (with \texttt{Pantheon+})\\
\hline
{\boldmath$10^{2}\omega{}_{\mathrm{b} }$} & $2.235\pm 0.011            $ & $2.237\pm 0.012            $ & $2.234\pm 0.012            $ & $2.238\pm 0.013            $   & $2.241\pm 0.012            $\\

{\boldmath$\omega{}_{\mathrm{cdm} }$} & $0.11965\pm 0.00099        $ & $0.1195\pm 0.0010          $ & $0.1198\pm 0.0012          $ & $0.1194\pm 0.0013          $   & $0.1190\pm 0.0013          $\\

{\boldmath$10^ 2\theta{}_{ \mathrm{s} }$} & $1.04205\pm 0.00025        $ & $1.04204\pm 0.00024        $ & $1.04202\pm 0.00025        $ & $1.04207\pm 0.00025        $   & $1.04207\pm 0.00025        $\\

{\boldmath$\tau{}_{\mathrm{reio} }$} & $0.0520\pm 0.0071          $ & $0.0480\pm 0.0087          $ & $0.0516\pm 0.0078          $ & $0.0483^{+0.0087}_{-0.0073}$    & $0.0485^{+0.0084}_{-0.0074}$\\

{\boldmath$w_{0}      $} & $-0.45^{+0.21}_{-0.23}     $ & $-0.47\pm 0.21             $ & $-0.44^{+0.23}_{-0.28}     $ & $-0.47^{+0.21}_{-0.25}     $    & $-0.960\pm 0.074           $\\

{\boldmath$w_{ \mathrm{a}}      $} & $-1.64^{+0.68}_{-0.57}     $ & $-1.56\pm 0.61             $ & $-1.65^{+0.84}_{-0.67}     $ & $-1.53^{+0.77}_{-0.58}     $  & $-0.15^{+0.28}_{-0.25}     $\\

{\boldmath$10^{9}A_{\mathrm{s} }$} & $2.101\pm 0.029            $ & $2.084^{+0.037}_{-0.032}   $ & $2.101\pm 0.032            $ & $2.086^{+0.037}_{-0.032}   $  & $2.083\pm 0.035            $\\

{\boldmath$n_{\mathrm{s} }  $} & $0.9698\pm 0.0033          $ & $0.9703\pm 0.0035          $ & $0.9691\pm 0.0038          $ & $0.9704\pm 0.0039          $  & $0.9710\pm 0.0040          $\\

{\boldmath$A_{\mathrm{L} }         $} & $1$ & $1.040^{+0.043}_{-0.048}   $ &    $1$ & $1.039\pm 0.046            $  & $1.070^{+0.045}_{-0.051}   $\\

{\boldmath$\Omega{}_{\mathrm{k} }  $} & $0$ &   $0$ & $0.0003\pm 0.0016          $ & $0.0001^{+0.0018}_{-0.0015}$   & $0.0015\pm 0.0013          $\\

\hline

{\boldmath$\Omega_{ \mathrm{DE},0}  $} & $0.650\pm 0.022            $ & $0.652\pm 0.021            $ & $0.648^{+0.026}_{-0.022}   $ & $0.652^{+0.024}_{-0.020}   $  & $0.6960\pm 0.0071          $\\

{\boldmath$H_0            $} & $63.9\pm 1.9               $ & $64.1\pm 1.8               $ & $63.9\pm 2.3               $ & $64.0\pm 2.1               $    & $68.56\pm 0.78             $\\

{\boldmath$S_{8 }         $} & $0.847\pm 0.016            $ & $0.841\pm 0.017            $ & $0.847\pm 0.016            $ & $0.840\pm 0.018            $    & $0.811^{+0.015}_{-0.014}   $\\
\hline

$\chi^ 2_{ \mathrm{min}}$ & 2489.05 & 2491.09 & 2489.41 & 2489.37 & 3004.69 \\

\hline

\end{tabular}
    \caption{Best-fits and 1$\sigma$ uncertainties of BI and PL with extended model of $A_{ \mathrm{L}}$ or $\Omega_{ \mathrm{k}}$ in \texttt{PSA} + \texttt{DESI DR2} or \texttt{PSA} + \texttt{DESI DR2} + \texttt{Pantheon+}. Some cases with fixed $A_{ \mathrm{L}} = 1$ or $\Omega_{ \mathrm{k}} = 0$ are showed as a fixed value without uncertainty. The last line of each table is the minimum value of $\chi^ 2$.}
    \label{tab:BI PL}
\end{table*}

\begin{figure*}
    \centering
    \includegraphics[width=0.9\linewidth]{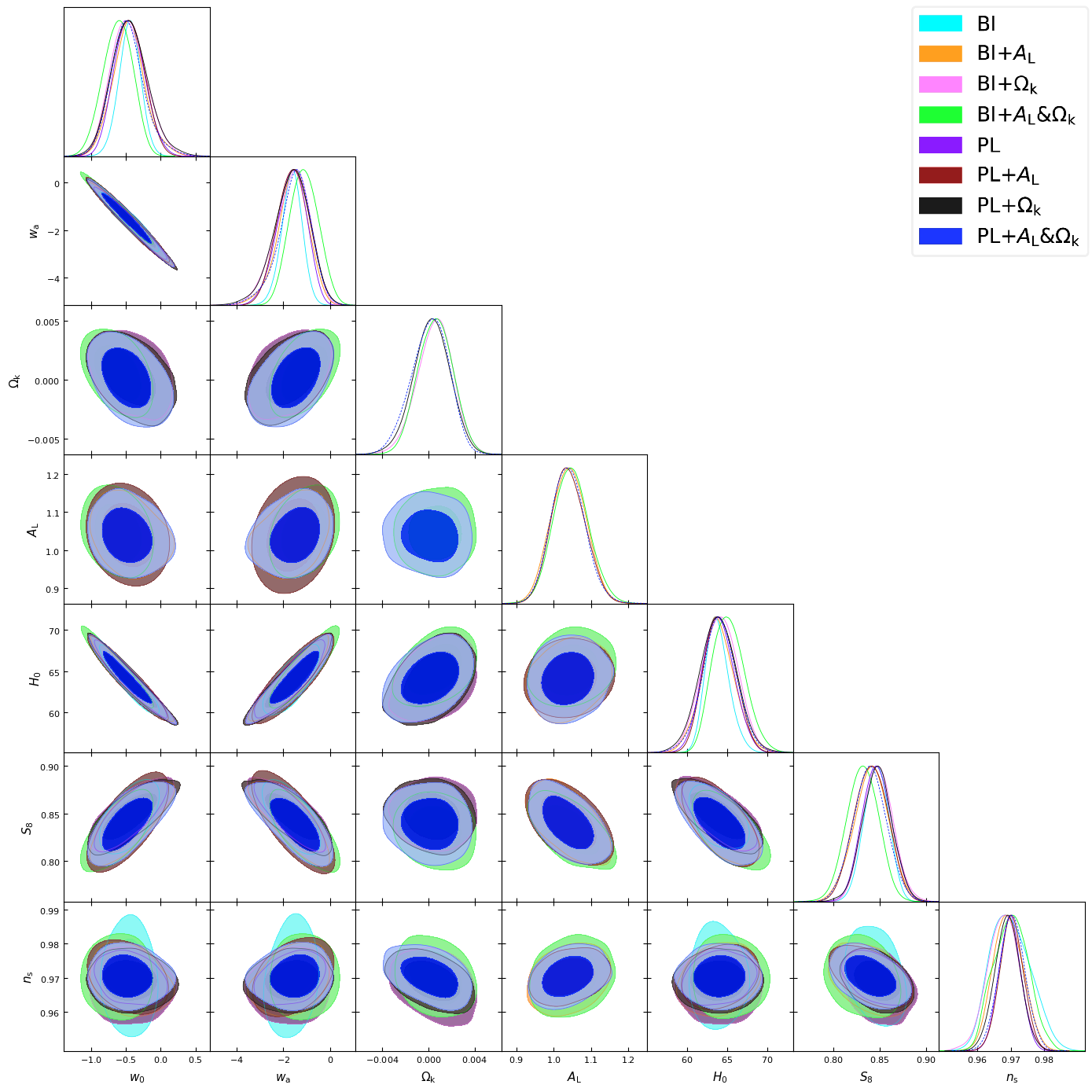}
    \caption{Contours of $w_ 0$, $w_{ \mathrm{a}}$, $\Omega_{ \mathrm{k}}$, $A_{ \mathrm{L}}$, $H_ 0$, $S_ 8$ and $n_{ \mathrm{s}}$ of BI and PL with \texttt{PSA} + \texttt{DESI DR2}. $d_{ \mathrm{M}}( z_ *)$ in BI and PL are similar, thus a less $w_ 0$ is consistent with a larger $H_ 0$ in BI with $A_{ \mathrm{L}}$ and $\Omega_{ \mathrm{k}}$.}
    \label{fig:BI PL CPL}
\end{figure*}
As previously noted, the BI scenario could reduce the Hubble tension to 3.2$\sigma$ with \texttt{Planck 2018} and also yields a larger $H_ 0$ in analyses involving \texttt{PSA} based on CPL parameterization or \texttt{PSA} + \texttt{DESI DR2} based on $\Lambda$CDM. However, with \texttt{DESI DR2}, \texttt{PSA} and CPL parameterization combined together, the best-fits $H_ 0$ for both cases are reduced to a lower value as approximately $64 \ \mathrm{km} \cdot \mathrm{s}^{ -1 } \cdot \mathrm{Mpc}^{ -1 }$. The best-fit value and 1$\sigma$ uncertainties for these cases are summarized in TAB. \ref{tab:BI PL}. From the table we can see that, for both cases of BI and PL, $\theta_ *$ is tightly constrained by \texttt{PSA} + \texttt{DESI DR2} under the CPL parameterization and takes similar values, while the degeneracy between $A_{ \mathrm{L}}$ and $n_{\mathrm{s}}$ as reported in \cite{Li:2025nnk} has a weaker impact on $H_ 0$ and $S_ 8$ compared to the influence of the CPL parameterization. Given nearly identical values of $\theta_ *$ and $\Omega_{ \mathrm{m}}$, the comoving angular diameter distance $d_{ \mathrm{M}}( z_ *)$ remains similar across models, preserving the $w_ 0$-$H_ 0$ degeneracy discussed in the previous section. 

Other than the combination of CPL dark energy effect and \texttt{DESI DR2}, in this section we also consider some extended models. We find in TAB. \ref{tab:BI PL} that with both $A_{ \mathrm{L}}$ and $\Omega_{ \mathrm{k}}$ free, the BI case labeled with ``BI + $A_{ \mathrm{L}}\&\Omega_{ \mathrm{k}}$'' yields the least $w_ 0$ as $w_ 0 = -0.61 \pm 0.21$, the highest $H_ 0$ as $H_ 0 = 65.2^{ + 1.8}_{ - 2.2} \ \mathrm{km} \cdot \mathrm{s}^{ -1 } \cdot \mathrm{Mpc}^{ -1 }$ and the lowest $S_ 8$ as $S_ 8 = 0.832 \pm 0.017$. On the other hand, the PL case labeled with ``PL + $A_{ \mathrm{L}}\&\Omega_{ \mathrm{k}}$'' yields $w_ 0 = -0.47^{ +0.21}_{ -0.25}$, $H_ 0 = 64.0 \pm 2.1 \ \mathrm{km} \cdot \mathrm{s}^{ -1 } \cdot \mathrm{Mpc}^{ -1 }$ and $S_ 8 = 0.840 \pm 0.018$. The marginalized 2D posterior contours for selected cosmological parameters are shown in FIG. \ref{fig:BI PL CPL}.

Both BI and PL yields positive value of $\Omega_{ \mathrm{k}}$, as $\Omega_{ \mathrm{k}} \approx 6 \times 10^{ - 4}$ for BI and $\Omega_{ \mathrm{k}} \approx 1 \times 10^{ -4}$ for PL with uncertainties on the order of $10^{ - 3}$, consistent with $\Omega_{ \mathrm{k}} = 0.0007 \pm 0.0019$ of Planck 2018 + BAO \cite{Planck:2018vyg}. All values of $\Omega_{ \mathrm{k}}$ are consistent with a spatially flat universe since they are all small enough within current observational uncertainties. Meanwhile, a non-vanishing spatial curvature also expands the selection range of $w_ 0$, but the very small deviation from zero does not affect $w_ 0$ too much. Spatial curvature affects the comoving angular diameter distance $d_{ \mathrm{M}}( z_ *)$ and thus contributes more directly than weak lensing amplitude $A_{ \mathrm{L}}$.

Moreover, the deviation between BI power spectrum and PL power spectrum as shown in \eqref{power spectrum of BI} and FIG. \ref{fig:BI power spectrum} can be compensated by adjustment in parameters such as weak lensing amplitude $A_{ \mathrm{L}}$ and CPL parametrization, therefore the best-fit values of $A_{ \mathrm{L}}$ in these two cases are different. As summarized in TAB. \ref{tab:BI PL}, $A_{ \mathrm{L}} = 1.040^{ + 0.043}_{ - 0.048}$ for ``PL + $A_{ \mathrm{L}}$'' and $A_{ \mathrm{L}} = 1.039 \pm 0.046$ for ``PL + $A_{ \mathrm{L}}\&\Omega_{ \mathrm{k}}$'', while $A_{ \mathrm{L}} = 1.040 \pm 0.047$ for ``BI + $A_{ \mathrm{L}}$'' and $A_{ \mathrm{L}} = 1.047 \pm 0.048$ for ``BI + $A_{ \mathrm{L}}\&\Omega_{ \mathrm{k}}$''. In the BI case, free $\Omega_{ \mathrm{k}}$ prefers larger $A_{ \mathrm{L}}$ values than its flat-universe counterparts. Furthermore, even though $A_{ \mathrm{L}}$ and $\Omega_{ \mathrm{k}}$ deviates only slightly from their theoretical value, the combination of a larger $A_{ \mathrm{L}}$ and a positive $\Omega_{ \mathrm{k}}$ in BI correlates with a less $w_ 0$ and a larger $w_{ a}$. Given the constraint on the comoving distance $d_{ \mathrm{M}}( z_ *)$ from \texttt{DESI DR2}, a less $w_ 0$ implies a larger $H_ 0$. However, since the inclusion of \texttt{DESI DR2} BAO data under the CPL parameterization tightens constraints on dark energy, reducing the allowed parameter space for $A_{ \mathrm{L}}$ and $\Omega_{ \mathrm{k}}$, which may suppress the apparent significance of oscillatory features in the BI power spectrum. Therefore, the aggravated tensions reported in DESI DR2 gets alleviated but remain statistically significant compared with the $\Lambda$CDM interpretation of CMB data.

\begin{figure}
    \centering
    \includegraphics[width=0.9\linewidth]{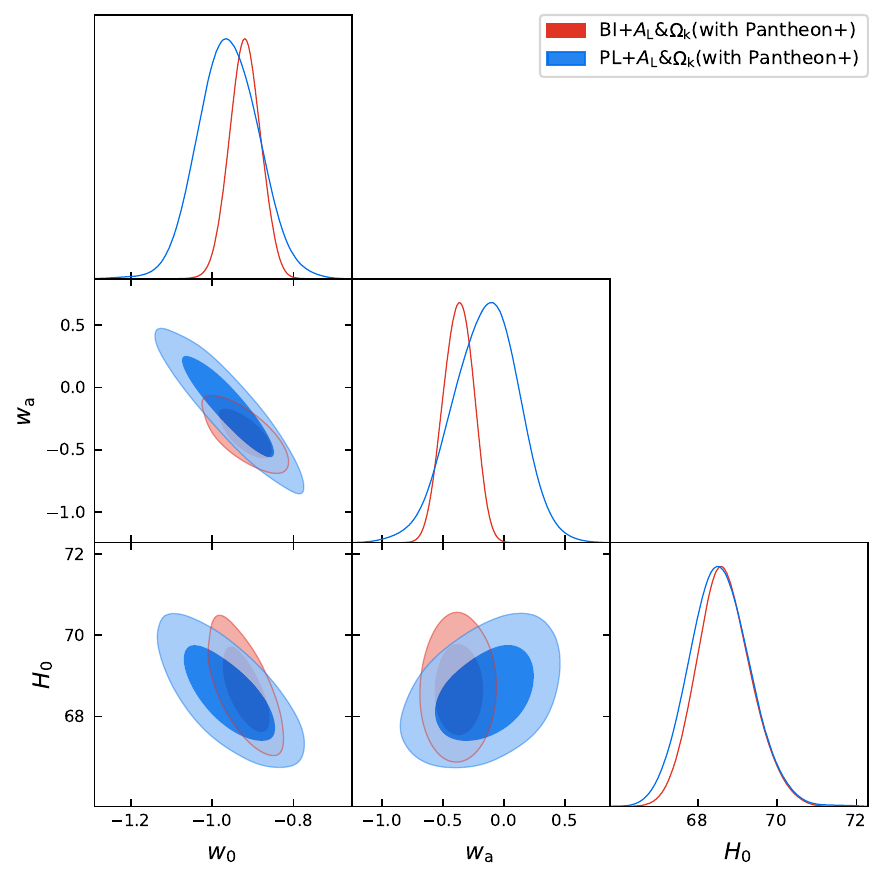}
    \caption{Contours of CPL dark energy of BI and PL with \texttt{PSA} + \texttt{DESI DR2} + \texttt{Pantheon+}. These two cases fit similar $H_0$, but the BI scenario prefers a evolving dark energy.}
    \label{fig:BI PL P}
\end{figure}

Additionally, we consider the observation of the SN Ia from \texttt{Pantheon+} as a supplement to \texttt{DESI DR2}. For flat-$w_ 0 w_{ \mathrm{a}}$CDM model, $w_ 0 = -0.851^{ +0.092}_{ -0.099}$ and $w_{ \mathrm{a}} = -0.70^{ +0.49}_{ -0.51}$ with \texttt{Planck \& Pantheon+}, $w_ 0 = -0.841^{ +0.066}_{ -0.061}$ and $w_{ \mathrm{a}} = -0.65^{ +0.28}_{ -0.32}$ with \texttt{Planck \& allBAO \& Pantheon+} \cite{Brout:2022vxf}. We show the combined constraints on PL and BI cases from \texttt{PSA} + \texttt{DESI DR2} + \texttt{Pantheon+} in FIG \ref{fig:BI PL P}. For the two cases, the most notable difference lies in the weak lensing amplitude $A_{ \mathrm{L}}$ and the dark energy parameters. The PL case gives $w_0 = -0.960 \pm 0.074$ and $w_{a} = -0.15^{ +0.28}_{ -0.25}$ with a larger $A_{ \mathrm{L}} = 1.070^{ +0.045}_{ -0.051}$, preferring $\Lambda$CDM in 1$\sigma$ confidence, while the BI case gives $w_ 0 = -0.919 \pm 0.038$ and $w_{ a} = -0.37 \pm 0.12$ with a less $A_{ \mathrm{L}} = 1.050^{ +0.035}_{ -0.039}$, generating 3$\sigma$ deviation from $\Lambda$CDM. In both case, the $H_0$ parameters has been raised up, with $H_0=68.66^{+0.63}_{-0.73}\ \mathrm{km} \cdot \mathrm{s}^{ -1 } \cdot \mathrm{Mpc}^{ -1 }$ for BI case, and $H_0=68.56\pm 0.78\ \mathrm{km} \cdot \mathrm{s}^{ -1 } \cdot \mathrm{Mpc}^{ -1 }$ for the PL case. Moreover, the PL case has a slightly less $S_ 8 = 0.811^{ +0.015}_{ -0.014}$ than the BI case $S_ 8 = 0.8234 \pm 0.0098$.

\section{Conclusion}
\label{Section: Conclusion}

In this paper, we perform numerical analysis of non-singular primordial bounce inflation scenario combined with CPL parameterization dark energy, and fit with the data from CMB (\texttt{PSA}), BAO (\texttt{DESI DR2}) and Supernova (\texttt{Pantheon+}). 

To investigate the effects of \texttt{DESI DR2} and the influence of CPL parameterization, we incorporate them individually in CMB analysis using \texttt{PSA} with the BI power spectrum or the PL power spectrum. We found that \texttt{DESI DR2} prefers a larger angular scale of BAO sound horizon in the framework of $\Lambda$CDM model. Thus, according to \eqref{sound horizon of BAO}, there is a tendency of a larger sound horizon of BAO $r_ *$ and a less comoving diameter distance from the observer to the recombination $d_ M( z_ *)$, which relates to a less matter parameter $\Omega_{ \mathrm{m}}$ and a larger energy density of dark energy $\rho_{ \mathrm{DE},0}$. Consequently, a larger Hubble constant $H_ 0 \approx 68.37 \ \mathrm{km} \cdot \mathrm{s}^{ -1 } \cdot \mathrm{Mpc}^{ -1 }$ is favored for both the BI scenario and PL power spectrum. On the other hand, CPL parameterization is weakly constrained by CMB data. Since $\theta_ *$, $\omega_{ \mathrm{b}}$ and $\omega_{ \mathrm{cdm}}$ are constrained tightly, the comoving angular diameter distance $d_ M(z)$ becomes more important. The calculation of $d_ M( z) = \int_ 0^ z dz ~ c / H( z)$ is primarily determined by the value of $H(z)$ at low redshift with a larger $H_ 0$ and large uncertainties. As shown in TAB. \ref{tab:DESI and CPL cosmo}, $w_ 0$ is fitted to a less value in ``BI no DESI'' than in ``PL no DESI'', which results in a larger $H_ 0$ as discussed in SEC. \ref{Section: The role of DESI DR2 and CPL}.

Considering the combination of \texttt{DESI DR2} and CPL parameterization dark energy, we employ the BI scenario and the PL power spectrum to investigate the cosmological parameters, with $A_{ \mathrm{L}}$ and $\Omega_{ \mathrm{k}}$ included as well. Under this consideration, several main conclusions are in order. (1) For both BI and PL cases, the dark energy parameter $w_ 0$ gets larger and the Hubble constant $H_0$ gets lowered, which aggravates the Hubble tension. (2) With $\Omega_k$ added as a free parameter, we get slightly positive $\Omega_k$ of ${\cal O}(10^{-4})$, consistent with the data of Planck 2018+BAO, which indicate a spatially flat universe. Since $\Omega_{ \mathrm{k}}$ affects the comoving angular diameter distance $d_{ \mathrm{M}}( z_ *)$ directly, $\Omega_{ \mathrm{k}}$ and $w_ 0$ are negative correlated.
(3) The weak lensing amplitude $A_{ \mathrm{L}}$ is slightly larger for the BI scenario and correlates with a less $w_ 0$ and a larger $w_{a}$. One has $A_{ \mathrm{L}}$ with best-fit value of $1.047$ for ``BI + $A_{ \mathrm{L}}\&\Omega_{ \mathrm{k}}$'' which is larger than both ``BI + $A_{ \mathrm{L}}$'' and ``PL + $A_{ \mathrm{L}}\&\Omega_{ \mathrm{k}}$'' cases. Moreover, $H_ 0 = 65.2^{ + 1.8}_{ - 2.2} \ \mathrm{km} \cdot \mathrm{s}^{ -1 } \cdot \mathrm{Mpc}^{ -1 }$ is given in ``BI + $A_{ \mathrm{L}}\&\Omega_{ \mathrm{k}}$'' as the largest value of all cases of BI or PL with \texttt{PSA} and \texttt{DESI DR2} within CPL parameterization. This is due to the less $w_0$ and the correlation between $w_0$ and $H_0$, and thus alleviated the Hubble tension, but only to a limited level due to the tight constraint of \texttt{DESI DR2} data on dark energy. (4) Finally, when adding \texttt{Pantheon+} into data, the value of $H_0$ for both ``BI + $A_{ \mathrm{L}}\&\Omega_{ \mathrm{k}}$'' and ``PL + $A_{ \mathrm{L}}\&\Omega_{ \mathrm{k}}$'' has risen up to approximately $68\ \mathrm{km} \cdot \mathrm{s}^{ -1 } \cdot \mathrm{Mpc}^{ -1 }$. The best-fit value for dark energy parameters $w_0$ and $w_a$ get raised as well. While the BI scenario has 3$\sigma$ deviation to $\Lambda$CDM, and the PL scenario is consistent with $\Lambda$CDM in 1$\sigma$ confidence level.

Along the line of this work, it is interesting to explore more deeply the theoretical constructions and the observational constraints on a complete theory of cosmic evolution, unifying the origin of the universe at the very beginning and the acceleration at the very end. With the accumulation of more and more precise observational data, one can have more clear understanding of the whole universe. Further discussions are left for the future work. 

\begin{acknowledgments}
We thank Yuxuan Li, Mengtian Jiang and Yifan Yang for useful discussions. This work is supported by the National Key Research and Development Program of China (Grant No. 2021YFC2203100) and the National Science Foundation of China (Grant No. 12575053). 
\end{acknowledgments}

\end{document}